\documentclass[10pt,twocolumn,letterpaper]{article}

\usepackage{cvpr}              % To produce the CAMERA-READY version
\definecolor{cvprblue}{rgb}{0.21,0.49,0.74}
\usepackage[pagebackref,breaklinks,colorlinks,allcolors=cvprblue]{hyperref}
\usepackage{algorithm}
\usepackage{algpseudocode}
\usepackage{amsthm}
\usepackage{amsmath}
\usepackage{amssymb}
\usepackage{bm}
\usepackage{xcolor}
\usepackage{multirow}
\usepackage{booktabs}
\usepackage{array}

\def\paperID{11990} % *** Enter the Paper ID here
\def\confName{CVPR}
\def\confYear{2025}

\title{Perception-based multiplicative noise removal using SDEs}

\author{An Vuong\\
Oregon State University\\
Corvallis, OR, 97331\\
{\tt\small vuonga2@oregonstate.edu}
\and
Thinh Nguyen\\
Oregon State University\\
Corvallis, OR, 97331\\
{\tt\small thinhq@oregonstate.edu}
}

\begin{document}
\maketitle
\begin{abstract}
Multiplicative noise, also known as speckle or pepper noise, commonly affects images produced by synthetic aperture radar (SAR), lasers, or optical lenses. Unlike additive noise, which typically arises from thermal processes or external factors, multiplicative noise is inherent to the system, originating from the fluctuation in diffuse reflections. These fluctuations result in multiple copies of the same signal with varying magnitudes being combined. Consequently, despeckling, or removing multiplicative noise, necessitates different techniques compared to those used for additive noise removal.

In this paper, we propose a novel approach using Stochastic Differential Equations based diffusion models to address multiplicative noise. We demonstrate that multiplicative noise can be effectively modeled as a Geometric Brownian Motion process in the logarithmic domain. Utilizing the Fokker-Planck equation, we derive the corresponding reverse process for image denoising. To validate our method, we conduct extensive experiments on two different datasets, comparing our approach to both classical signal processing techniques and contemporary CNN-based noise removal models. Our results indicate that the proposed method significantly outperforms existing methods on perception-based metrics such as FID and LPIPS, while maintaining competitive performance on traditional metrics like PSNR and SSIM. 

% Code and appendix available at \url{https://github.com/anvuongb/sde_multiplicative_noise_removal}.
\end{abstract}    
\section{Introduction}
\label{sec:intro}

Multiplicative noise removal is a long standing problem in computer vision and has been studied by many researchers over the past few decades \cite{huang2009new}\cite{bioucas2010multiplicative}\cite{huang2012multiplicative}\cite{shan2019multiplicative}\cite{feng2021models}. Unlike additive noise, which is usually the result of thermal fluctuations during image acquisition or transmission, multiplicative noise happens when multiple copies of the signal with random scaling factors are added together. This often happens due to the internal physical construction of the image capturing devices, i.e. optical lenses, radar/laser imaging, ultrasound sensors, etc. Because of this, removing multiplicative noise, sometimes referred to as despeckling, often requires more sophisticated approaches compared to its counterpart additive noise. Popular approaches include modelling the noise using Partial Differential Equations (PDEs) \cite{yu2002speckle} \cite{chen2012multiplicative}, converting into additive domain and optimize using Total Variation (TV) objective \cite{shi2008nonlinear}, and applying MAP estimation \cite{aubert2008variational}. Classical methods based on block-matching technique also works decently for this problem \cite{dabov2007image}.
\begin{figure}\small
    \setlength{\tabcolsep}{1.5pt}
    \centering
    \begin{tabular*}{0.35\textwidth}{@{\extracolsep{\fill}} ccc }
    Original & Noisy & Denoised
    \end{tabular*}
  \centering
  \includegraphics[width=0.45\textwidth]{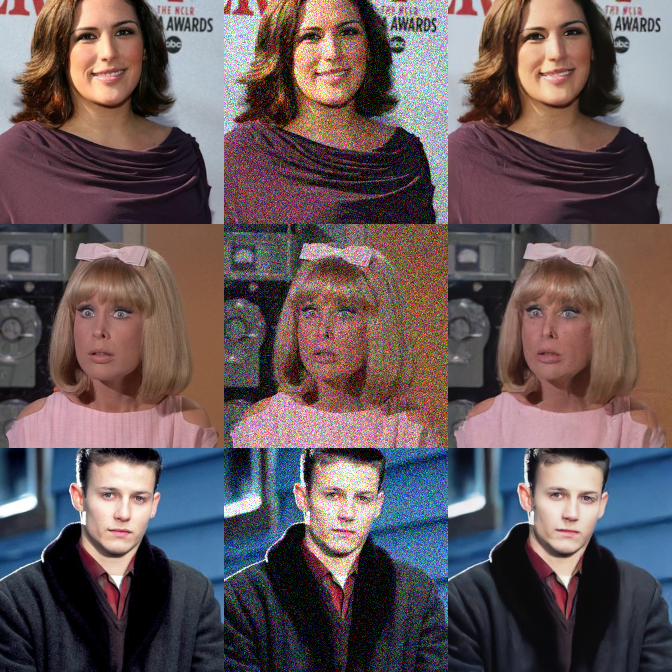}
  \caption{\small Samples generated by our methods, on images randomly selected from CelebA dataset. From left to right are the original, corrupted by multiplicative noise (noise level $0.08$), and denoised versions.}
\end{figure}
More recently, deep learning based methods have been introduced with great successes in denoising performance \cite{zhang2017dncnn}\cite{basicsr}\cite{chen2022simple}. These methods usually use image-to-image translation architecture, where the neural networks directly predict the clean images, or the amount of noise generated by the stochastic process, without much assumption on the noise dynamics. Thus, many of these models can be applied to reverse different kinds of corruptions, including multiplicative noise. However, these techniques mostly rely on "per-pixel" metrics such as MSE, PSNR, or SSIM, which has been observed to not correlate well with human perception \cite{blau2018perception}.

In this work, we propose the novel application of Stochastic Differential Equations (SDEs) to perform multiplicative noise removal. We show that the dynamics of multiplicative noise is well captured by SDEs, specifically Geometric Brownian motion. We then derive the reverse SDEs which are used to generate denoised samples. By running extensive experiments on two different datasets, we demonstrate the effectiveness of our method on creating clean images that achieve high perception scores. We detail the construction of our approach in Section \ref{sec:methods}.

\section{Related works}
Over the last few years, with the advances of deep learning, there has been active research in applying CNN to image denoising problems, especially for speckle or multiplicative noise. Notable works can be found in \cite{li2021speckle}\cite{choi2018speckle}\cite{feng2020speckle}. In these texts, the common theme is to perform image-to-image translation with a Convolutional neural network (CNN) acting as the mapping function. This CNN is usually trained to minimize MSE or PSNR loss directly on the pairs of clean and noisy images. Some works propose to use specially-crafted features as the input, such as frequency features \cite{yang2022fddl}, or wavelet features \cite{zhang2019sar}, and sub-bands \cite{liu2021contourlet}. These works are usually limited to grayscale images, and are often matched in performance by DnCNN\cite{zhang2017dncnn}, and can be beaten by NAFNet\cite{chen2022simple}, MPRNet\cite{Zamir2021MPRNet}, or Restormer\cite{Zamir2021Restormer}. 

Recently, there is a line of works applying diffusion technique to this problem. In \cite{guha2023sddpm}, the authors propose a DDPM-like architecture for despeckling, but this model needs to be re-trained for each noise level. Similarly, \cite{perera2023sar}\cite{xiao2023unsupervised} also use DDPM framework with minor modifications. These works still limit their testing to greyscale images only. We find the discussion in \cite{pearl2023svnr} to be the most related to our work, albeit with different assumption of the noise characteristics and the construction of the diffusion process, where the authors still rely on the DDPM equations.

To the best of our knowledge, we are the first to directly model this problem using SDE, which captures the dynamics of the noise process, and derive the sampling equation which is then used to perform denoising.
\section{Methods}
\label{sec:methods}

\subsection{SDE and Diffusion models}

In this section, we give a brief overview of Stochastic Differential Equations, Itô's calculus, and the application to generative modelling.

Let $\beta(t)$ be a Brownian motion indexed by time $t$, i.e. $\beta(t)$ is a random process with independent and zero-mean Gaussian increment, then the classic result from Itô's calculus gives
\begin{equation}
    \int \beta(t)d\beta(t) = \frac{1}{2}\beta^{2}(t) -\frac{1}{2}t
\end{equation} or in a more compact differential form
\begin{equation}
    \frac{1}{2}d\beta^{2}(t)= \beta(t)d\beta(t) +\frac{1}{2}dt\label{eq:ito_formula1}
\end{equation}
This seminal result plays a central role in solving SDE, whose informal definition can be given by the following differential equation
\begin{equation}
    dx(t) = f(x(t), t)dt + L(x(t), t)d\beta(t) \label{eq:sde_def}
\end{equation}
where $f(.)$, $L(.)$ are some functions,  $x(t)$ is the random process of interest, and $dx(t)$ represents the (random) infinitesimal change of $x(t)$. The remarkable thing about \eqref{eq:sde_def} is that, under some smoothness assumptions of $f(.)$ and $L(.)$, there exists a unique SDE that models the reverse process \cite{oksendal2003stochastic}:
\begin{align}
    dx(T-t) = &-f(x(T-t), T-t)dt \nonumber\\
    & + L(x(T-t), T-t)d\beta(T-t)  \label{eq:sde_reverse}\\
    &+ L^{2}(T-t)\nabla\log p_{T-t}dt \nonumber
\end{align}
where $p_{T-t}$ denotes the distribution of $x(T-t)$. This result comes directly from the application of the Fokker-Planck equation, and is sometimes referred to as Anderson's theorem. The extension of these results to multivariate process is straightforward, and detailed proofs can be found in \cite{ito1951stochastic}\cite{oksendal2003stochastic}\cite{sarkka2019applied}.

A crucial observation is that, if $\nabla \log p_{T-t}$ is known, then one can can run equation \eqref{eq:sde_reverse} to generate new sample $x({T-t})$ that comes from data distribution $p_{T-t}(x)$. This motivates the search for an efficient method to estimate $\nabla \log p_{T-t}$, also known as the score function. Let us now denote the estimation of $\nabla \log p(x)$ as $s_{\bm{\theta}}(x)$, some function parameterized by $\bm{\theta}$, then the surprising result from \cite{hyvarinen2005estimation} gives
\begin{align}
    \mathbb{E}_{x}\Big[\frac{1}{2}||& \nabla\log p(x) -  s_{\bm{\theta}}(x)||^{2}_{2}\Big] \nonumber\\
    &\propto \mathbb{E}_{x}\Big[\frac{1}{2}||   s_{\bm{\theta}}(x)||^{2}_{2}+\textrm{trace}\big(\nabla\mathbf{s}_{\bm{\theta}}(x)\big)\Big] \label{eq:score_matching}
\end{align}
From \eqref{eq:score_matching}, one can parameterize $s_{\bm{\theta}}(x)$ as a neural network and directly optimize using samples of $x$ without needing the knowledge of $\nabla \log p(x)$. This result, along with Langevin sampling, was used in \cite{song2019generative} to train neural networks for images generation tasks that achieve high-quality results. In the follow-up papers \cite{song2020denoising}\cite{song2019generative}, the authors tweaked the learning procedure to learn $\nabla \log p_{T-t}(x)$, instead of $\nabla \log p(x)$, this proved to be more stable and easier to implement, while achieving better results at the same time. In these works, the sampling was done using \eqref{eq:sde_reverse}, instead of Langevin dynamics. This approach is referred to as Score-based Generative Models (SGMs).

In a parallel development, \cite{ho2020denoising} proposed a similar framework from Markov chain perspective, which is named Diffusion Denoising Probabilistic Models (DDPM). DDPM formulation leads to a much simpler objective function, which is shown to be equivalent to \eqref{eq:score_matching} under Gaussian noise assumption \cite{vincent2011connection}. In this paper, we use the SDE-based formulation since it is more flexible, allowing us to directly model the desired underlying dynamics of the noise process.

\subsection{Noise models}
In this section, we introduce multiplicative noise, and show how SDE can be used to model the dynamics of this process.

A real-valued signal $x \in \mathbb{R}$ is corrupted by multiplicative noise is modeled as 
\begin{equation}
    \Tilde{x} = \epsilon x
\end{equation}
where $\epsilon \in \mathbb{R}$ is a random variable, usually modeled as having Gamma or Log-normal distribution \cite{arsenault1976properties}, and $\Tilde{x}$ is the corrupted version of $x$. Here, we extend this noise process to multi-dimensional $\mathbf{x}\in \mathbb{R}^{d}$, with the assumption that this corruption affects each component independently
\begin{equation}
    \Tilde{\mathbf{x}} = \mathbf{x}\odot\bm{\epsilon} \label{eq:mult_noise_model}
\end{equation}
where $\bm{\epsilon} \in \mathbb{R}^{d}$ and $\odot$ represents the element-wise multiplication. We now show that \eqref{eq:mult_noise_model} can be well modelled by the following SDE
\begin{equation}
    d\mathbf{x}=\alpha(t)\mathbf{x}(t)\odot d\bm{\beta}(t) \label{eq:mult_noise_sde}
\end{equation}
where $\alpha(t)$ is some time-varying scalar function and $\bm{\beta}(t)$ is a Brownian motion on $\mathbb{R}^{d}$. Indeed, the solution to \eqref{eq:mult_noise_sde} (details in the Appendix\footnote{Appendix is provided in Supplemental materials}) is given as
\begin{align}
    x_{t,i} &= x_{0,i}\exp\Big(\int_{0}^{t}\frac{1}{2}\alpha^{2}(\tau)d\tau + \int_{0}^{t}\alpha(\tau)d\beta(\tau)\Big)\label{eq:mult_noise}\\
    &=x_{0,i}\exp\Big( -\int_{0}^{t}\frac{1}{2}\alpha^{2}(\tau)d\tau + \Big(\int_{0}^{t}\alpha^{2}(\tau)d\tau\Big)^{\frac{1}{2}}n \Big)\\
    &\quad\quad\quad\quad n\sim\mathcal{N}(0,1) \nonumber
\end{align}
where $x_{t,i}$ denotes the $i$-th entry of $\mathbf{x}(t)$. Since $n$ is Gaussian, the exponential term in \eqref{eq:mult_noise_sde} will follow Log-normal distribution, satisfying our previous assumption on $\epsilon$. If we select $\mathbf{x}(0)$ to be the clean image $\mathbf{x}$, then with appropriate value of $t$, $\Tilde{\mathbf{x}}$ can be well modelled by \eqref{eq:mult_noise_sde}. 

We can now apply Anderson's theorem to derive the reverse SDE for \eqref{eq:mult_noise_sde} and use score-matching method from \cite{song2020score} to construct a denoising model. But this formulation gives a rather complicated reverse SDE, as shown in the Appendix. For this reason, we propose to apply a simple logarithmic transformation to $\mathbf{x}$, this yields a much simpler reverse SDE, with the additional advantage of being able to apply the results from \cite{vincent2011connection}, making the loss function easier to derive.

\subsection{Loss function in the logarithmic domain}

Let us denote $y_{t, i}=\log x_{t, i}$. Now, equation \eqref{eq:mult_noise} becomes
\begin{align}
    y_{t,i} = y_{0,i} -\int_{0}^{t}\frac{1}{2}\alpha^{2}(\tau)d\tau + \int_{0}^{t}\alpha(\tau)d\beta(\tau)
\end{align}
which can also be expressed in differential form to obtain the SDE
\begin{equation}
     dy_{t,i}= -\frac{1}{2}\alpha^{2}(t)dt + \alpha(t)d\beta(t)\label{eq:mult_noise_log_sde}
\end{equation}
% This can be shown by simple substitution
% \begin{align}
%     dy_{t,i} &= d\Big( y_{0,i} -\int_{0}^{t}\frac{1}{2}\alpha^{2}(\tau)d\tau + \int_{0}^{t}\alpha(\tau)d\beta(\tau)\Big) \nonumber \\
%     &= 0-\frac{1}{2}\alpha^{2}(t)dt + \alpha(t)d\beta(t)\nonumber\\
%     &= -\frac{1}{2}\alpha^{2}(t)dt + \alpha(t)d\beta(t)\nonumber
% \end{align}
% where the last line made use of the fact that Brownian motion can be seen as weak derivative of white noise \cite{sarkka2019applied}.

Equation \eqref{eq:mult_noise_log_sde} can also be written in vector form
\begin{equation}
    d\mathbf{y}_{t} = -\frac{1}{2}\alpha^{2}(t)\mathbf{1}dt + \alpha(t)d\bm{\beta}(t)\label{eq:mult_noise_log_vec_sde}
\end{equation}
which has the corresponding reverse SDE (proof in the Appendix)
\begin{align}
    d\mathbf{y}_{T-t} = \Big(\frac{1}{2}\alpha^{2}(T-t)\mathbf{1} &+ \alpha^{2}(T-t)\nabla\log p_{T-t}(\mathbf{y}_{T-t})\Big)dt \nonumber\\
    &+ \alpha(T-t)d\bm{\beta}(T-t) \label{eq:mult_noise_log_vec_reverse_sde}
\end{align}
where $T$ is the terminal time index, i.e. at which the forward SDE \eqref{eq:mult_noise_log_vec_sde} stopped. 

Applying Euler-Maruyama discretization to \eqref{eq:mult_noise_log_vec_sde} and \eqref{eq:mult_noise_log_vec_reverse_sde}, where $\alpha(t)$ is selected to be $\sqrt{\frac{d\sigma(t)}{dt}}$ with $\sigma(t)$ is some differentiable function having non-negative slope, gives the following pair of SDEs
\begin{align}
    \mathbf{y}_{k} &= \mathbf{y}_{k-1} -\frac{1}{2}\big(\sigma(k)-\sigma(k-1)\big)\mathbf{1} \nonumber\\
    &\;\;\;\;\;\;\;\;\;\;\;\;+ \sqrt{\sigma(k)-\sigma(k-1)}\mathbf{n}_{k}\nonumber\\
    &= \mathbf{y}_{0} -\frac{1}{2}\big(\sigma(k)-\sigma(0)\big)\mathbf{1} + \sqrt{\sigma(k)-\sigma(0)}\mathbf{n}_{k}\label{eq:discrete_forward}\\
    \mathbf{y}_{K-k} &=\mathbf{y}_{K-k+1} \nonumber\\
    &+ \frac{1}{2}\Big(\sigma(K-k+1)-\sigma(K-k)\Big)\Big( \mathbf{1} + \nonumber\\
    &\;\;\;\;\;\;\;\;\;\;\;\;\;\;2 \nabla\log p_{K-k+1}(\mathbf{y}_{K-k+1})\Big) \nonumber\\
    &+ \sqrt{\sigma(K-k+1)-\sigma(K-k)} \mathbf{n}_{k}\label{eq:discrete_reverse}\\
    &\;\;\;\;\;\;\;\;\;\;\;\;\mathbf{n}_{k}\sim\mathcal{N}(\mathbf{0},\mathbf{1})\nonumber
\end{align}
To derive the loss function, note that equation \eqref{eq:discrete_forward} has a Gaussian transition kernel $p(\mathbf{y}_{k}|\mathbf{y}_{k-1}) = \mathcal{N}\Big(\mathbf{y}_{k}; \mathbf{y}_{k-1} -\frac{1}{2}\big(\sigma(k)-\sigma(k-1)\big)\mathbf{1}, \sigma(k)-\sigma(k-1)\Big)$. Thus, results from \cite{vincent2011connection} applies, which states the following connection between SGMs and DDPMs: let $\mathbf{s}^{*
}(\mathbf{x}_{t}, t)$ and $\mathbf{n}^{*}(\mathbf{x}_{t}, t)$ be the minimizers of SGM and DDPM objectives, respectively, then $\mathbf{n}^{*}(\mathbf{x}_{t}, t)=-\sqrt{\textrm{var}(\mathbf{x}_{t})}\mathbf{s}^{*
}(\mathbf{x}_{t}, t)$ if $p(\mathbf{x}_{t}|\mathbf{x}_{t-1})$ is Gaussian, where $\textrm{var}(\mathbf{x}_{t})$ denotes the variance of the stochastic process at time $t$. This means the denoising objective from DDPM can be readily applied to our formulation in the logarithmic domain, giving the following trainable loss
\begin{align}
    \mathcal{L}_{\bm{\theta}, \textrm{discete}} = \mathbb{E}_{\mathbf{y}}\mathbb{E}_{k}\Big[|| \bm{n}_{k} + \sqrt{\sigma(k)-\sigma(0)}\mathbf{s}_{\bm{\theta}}( \mathbf{y}_{k},k)||^{2}_{2}\Big]\label{eq:sde_loss}
\end{align}
where the plus sign is due to the score function being opposite in direction to the noise vector.
\begin{figure}\small
    \setlength{\tabcolsep}{1.5pt}
    \begin{tabular*}{0.43\textwidth}{@{\extracolsep{\fill}} ccccc }
    Original & Noisy & ODE & DDIM & Stochastic
    \end{tabular*}
  \centering
  \includegraphics[width=0.45\textwidth]{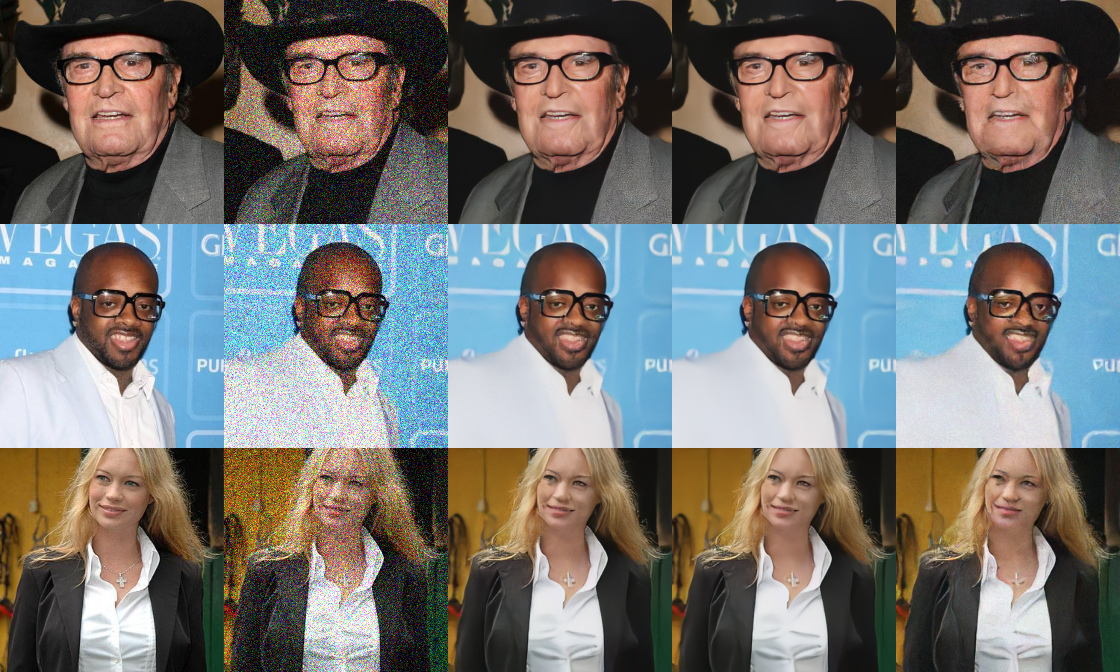}
  \caption{\small Comparing between different sampling techniques on randomly selected CelebA images, at noise level $0.12$. The first two columns include the original images and their noised versions, respectively. These are followed by the results generated by our model using ODE, DDIM, and stochastic samplers, respectively. High resolution version is available in the appendix. }\label{fig:quali_sampling_methods}
\end{figure}
Thus, by transforming the data into the logarithmic domain, we can take advantage of the simplicity of DDPM framework, while still being able to accurately capture our noise model thanks to the flexibility of SDE. Furthermore, since $\log$ is a bijective transformation for positive variables, we can easily recover our images from the samples generated in logarithmic domain by taking the exponential. The training and sampling procedures are summarized in Algorithms \ref{alg:training} and \ref{alg:sampling_stochastic}.

\begin{algorithm}
    \caption{Training procedure of $\mathbf{s}_{\bm{\theta}}(\mathbf{y}_k, k)$}\label{alg:training}
    \textbf{Input} Training data $\mathbf{x}\in\mathcal{X}$, noise schedule $\sigma(t)$, number of epochs $E$, total steps $K$\\
    \textbf{Output} Score network $\mathbf{s}_{\bm{\theta}}(\mathbf{y}_k, k)$, where $\mathbf{y}=\log\mathbf{x}$
    \begin{algorithmic}[1]
    \State $i = 1$
    \While{$i \leq E$}
        \For{$\mathbf{x}\in\mathcal{X}$}
            \State $\mathbf{y}=\log\mathbf{x}$
            \State Draw $k\sim\mathcal{U}(0,K)$
            \State Calculate $\mathbf{y}_{k}$ according to \eqref{eq:discrete_forward}
            \State Take gradient step on:
            \Statex $\quad\quad\quad\quad\mathbb{E}_{\mathbf{y}}\mathbb{E}_{k}\Big[|| \bm{n}_{k} + \sqrt{\sigma(k)-\sigma(0)}\mathbf{s}_{\bm{\theta}}( \mathbf{y}_{k},k)||^{2}_{2}\Big]$
        \EndFor
        \State $i = i+1$
    \EndWhile
    \State Return $\mathbf{s}_{\bm{\theta}}(\mathbf{y}_{k},k)$
    \end{algorithmic}
\end{algorithm}

\begin{algorithm}
    \caption{Noise removal procedure (stochastic)}\label{alg:sampling_stochastic}
    \textbf{Input} Corrupted image $\Tilde{\mathbf{x}}$, number of denoising steps $K$\\
    \textbf{Output} Denoised image $\hat{\mathbf{x}}$
    \begin{algorithmic}[1]
    \State $k=1$, $\mathbf{y}_{K} = \log\Tilde{\mathbf{x}}$
    \While{$k \leq K$}
        \State $\nabla \log p_{K-k+1}=\mathbf{s}_{\bm{\theta}}(\mathbf{y}_{K-k+1}, K-k+1)$
        \State Sample $\mathbf{y}_{K-k}$ using \eqref{eq:discrete_reverse} 
        % with $\mathbf{x}_{K-k+1}\odot\nabla\log p_{K-k+1}(\mathbf{x}_{K-k+1}) = \mathbf{s}_{K-k+1}-1$
    \EndWhile
    \State Return denoised image $\hat{\mathbf{x}}=\exp(\mathbf{y}_{0})$
    \end{algorithmic}
\end{algorithm}

\subsection{Enhancing image quality by deterministic sampling}
Since the sampling equation \eqref{eq:mult_noise_log_vec_reverse_sde} is a stochastic process, given a noisy input, different runs will result in different denoised images, albeit with the same statistics. This might not be desirable in applications where determinism is required. Thus, we now give two deterministic sampling equations specifically for our proposed SDE. Experimentally, we show that samples generated deterministically achieve substantially better Fréchet inception distance (FID) \cite{heusel2017gans} and Learned Perceptual Image Patch Similarity (LPIPS) \cite{zhang2018perceptual} metrics

\textbf{Probability flows}. Following the idea in \cite{song2020score}, from \eqref{eq:mult_noise_log_vec_reverse_sde}, it can be shown that there exists an ODE counterpart whose marginals match that of \eqref{eq:mult_noise_log_vec_reverse_sde}, this ODE is given as
\small
\begin{align}
    d\mathbf{y}_{T-t} = \Big(\frac{1}{2}\alpha^{2}(T-t)\mathbf{1} &+ \frac{1}{2}\alpha^{2}(T-t)\nabla\log p_{T-t}(\mathbf{y}_{T-t})\Big)dt \label{eq:mult_noise_log_vec_reverse_ode}
\end{align}
\normalsize
which yields the disrecte-time equation
\begin{align}
    \mathbf{y}_{K-k} &=\mathbf{y}_{K-k+1} \nonumber\\
    &+ \frac{1}{2}\Big(\sigma(K-k+1)-\sigma(K-k)\Big)\Big( \mathbf{1} + \nonumber\\
    &\;\;\;\;\;\;\;\;\;\;\;\;\;\;\nabla\log p_{K-k+1}(\mathbf{y}_{K-k+1})\Big) \label{eq:discrete_ode_reverse}
\end{align}

\textbf{Implicit probabilistic models}. From the works of \cite{mohamed2016learning} \cite{song2020denoising}, given a stochastic process with Markov transition kernel $p(\mathbf{z}_{t-1}|\mathbf{z}_{t})$, one can design another process with non-Markov transition kernel $q(\mathbf{z}_{t-1}|\mathbf{z}_{t}, \mathbf{z}_{0})$ whose marginals match, i.e $q(\mathbf{z}_{t-1}|\mathbf{z}_{t}) = p(\mathbf{z}_{t-1}|\mathbf{z}_{t})$. For our discretized SDE in \eqref{eq:discrete_forward}, its non-Markov counterpart is given as (derivations can be found in Appendix)
\begin{align}
    q&(\mathbf{y}_{k-1}|\mathbf{y}_{k}, \mathbf{y}_{0}) = \mathcal{N}(\bm{\mu}_{k}, \bm{\Sigma}_{k})\\
    &\bm{\mu}_{k}= \mathbf{y}_{0}-\frac{1}{2}\eta(k-1)\mathbf{1} \nonumber\\
    &\;\;\;\;\;\;\;\;+ \frac{\sqrt{\eta(k-1)-\zeta_{k}^{2}}}{\sqrt{\eta(k)}}\big(\mathbf{y}_{k}-\mathbf{y}_{0}+\frac{1}{2}\eta(k)\mathbf{1}\big)\\
    &\bm{\Sigma}_{k}=\zeta_{k}^{2}\mathbf{I}
\end{align}
where $\zeta_{k}^{2}$ is a new parameter controlling the variance of the process, and $\eta(k)=\sigma(k)-\sigma(0)$. For deterministic sampling, we can set $\zeta_{k}^{2}=0$ and obtain the sampling equations
\begin{align}
    \mathbf{y}&_{k-1} = \hat{\mathbf{{y}}}_{0} -\frac{1}{2}\eta(k-1)\mathbf{1} \nonumber\\
    &\;\;\;\;\;\;\;\;+ \frac{\sqrt{\eta(k-1)}}{\sqrt{\eta(k)}}\big(\mathbf{y}_{k}-\hat{\mathbf{y}}_{0}+\frac{1}{2}\eta(k)\mathbf{1}\big) \label{eq:ddim_sampling}
\end{align}
and $\hat{\mathbf{{y}}}_{0}$ is given as
\begin{equation}
    \hat{\mathbf{{y}}}_{0} = \mathbf{y}_{k} + \frac{1}{2}\eta(k)\mathbf{1} + \eta(k)\mathbf{s}_{\bm{\theta}}( \mathbf{y}_{k},k) \label{eq:ddim_y0}
\end{equation}
where $\mathbf{s}_{\bm{\theta}}( \mathbf{y}_{k},k)$ is the output of the neural network defined by the objective \eqref{eq:sde_loss}. Equation \eqref{eq:ddim_y0} can be thought of as the best possible prediction of $\mathbf{y}_{0}$ in single-step using $\mathbf{y}_{k}$. We note that this approach defines an implicit model that matches the marginals of \eqref{eq:discrete_forward}, so the trained $\mathbf{s}_{\bm{\theta}}( \mathbf{y}_{k},k)$ can be reused even though we make no use of the original SDE during sampling. This sampler is referred to as Denoising Diffusion Implicit Models (DDIMs) in the literature.

Sampling procedure for these two techniques are provided in Algorithm \ref{alg:sampling_ode} and \ref{alg:sampling_ddim}.

\begin{algorithm}
    \caption{Noise removal procedure (ODE)}\label{alg:sampling_ode}
    \textbf{Input} Corrupted image $\Tilde{\mathbf{x}}$, number of denoising steps $K$\\
    \textbf{Output} Denoised image $\hat{\mathbf{x}}$
    \begin{algorithmic}[1]
    \State $k=1$, $\mathbf{y}_{K} = \log\Tilde{\mathbf{x}}$
    \While{$k \leq K$}
        \State $\nabla \log p_{K-k+1}=\mathbf{s}_{\bm{\theta}}(\mathbf{y}_{K-k+1}, K-k+1)$
        \State Compute $\mathbf{y}_{K-k}$ using \eqref{eq:discrete_ode_reverse} 
        \State $k=k+1$
    \EndWhile
    \State Return denoised image $\hat{\mathbf{x}}=\exp(\mathbf{y}_{0})$
    \end{algorithmic}
\end{algorithm}
\begin{algorithm}
    \caption{Noise removal procedure (DDIM)}\label{alg:sampling_ddim}
    \textbf{Input} Corrupted image $\Tilde{\mathbf{x}}$, number of denoising steps $K$\\
    \textbf{Output} Denoised image $\hat{\mathbf{x}}$
    \begin{algorithmic}[1]
    \State $k=K$, $\mathbf{y}_{K} = \log\Tilde{\mathbf{x}}$
    \While{$k \leq K$}
        \State Calculate 
        \State $\nabla \log p_{k}=\mathbf{s}_{\bm{\theta}}(\mathbf{y}_{k}, k)$
        \State Compute $\hat{\mathbf{{y}}}_{0}$ using \eqref{eq:ddim_y0}
        \State Compute $\mathbf{y}_{k-1}$ using \eqref{eq:ddim_sampling}
        \State $k=k-1$
    \EndWhile
    \State Return denoised image $\hat{\mathbf{x}}=\exp(\mathbf{y}_{0})$
    \end{algorithmic}
\end{algorithm}
\section{Experiments}
\begin{table*}
\centering
\begin{tabular}{|c|>{\centering\arraybackslash}m{1.5cm}|>{\centering\arraybackslash}m{1.5cm}|>{\centering\arraybackslash}m{1.5cm}|>{\centering\arraybackslash}m{1.5cm}|>{\centering\arraybackslash}m{1.5cm}|}
\hline
\textbf{Sampling technique} & \textbf{FID $\downarrow$} & \textbf{LPIPS $\downarrow$} & \textbf{PSNR $\uparrow$} & \textbf{SSIM $\uparrow$} \\ \hline\hline
 ODE & \textbf{13.9156} & \textbf{0.0365} & \textbf{31.8902} & \textbf{0.9348} \\ 
  DDIM & 25.3188	& 0.0882 & 28.6549 & 0.9032 \\ 
  Stochastic & 32.3811 & 0.1075 & 26.8267 & 0.8493 \\ 
\hline
\end{tabular}
\caption{Comparison of different sampling techniques on CelebA dataset at noise level $0.12$ }
\label{tab:sampling_celeba}
\end{table*}
\begin{table*}
\centering
\resizebox{\linewidth}{!}{%
\begin{tabular}{|c|>{\centering\arraybackslash}m{2cm}||>{\centering\arraybackslash}m{1.5cm}|>{\centering\arraybackslash}m{1.5cm}|>{\centering\arraybackslash}m{1.5cm}|>{\centering\arraybackslash}m{1.5cm}||>{\centering\arraybackslash}m{1.5cm}|>{\centering\arraybackslash}m{1.5cm}|>{\centering\arraybackslash}m{1.5cm}|>{\centering\arraybackslash}m{1.5cm}|}
\hline
\multicolumn{2}{|c||}{\textbf{Dataset}} &  \multicolumn{4}{c||}{\textbf{CelebA}} &  \multicolumn{4}{c|}{\textbf{UC Merced Landuse}} \\
\hline
\textbf{Noise level} & \textbf{Model} & \textbf{FID $\downarrow$} & \textbf{LPIPS $\downarrow$} & \textbf{PSNR $\uparrow$} & \textbf{SSIM $\uparrow$} & \textbf{FID $\downarrow$} & \textbf{LPIPS $\downarrow$} & \textbf{PSNR $\uparrow$} & \textbf{SSIM $\uparrow$} \\ 
\hline\hline
\multirow{5}{*}{0.04} & Ours (ODE) & \textbf{13.9156} & \textbf{0.0365} & 31.8902 & 0.9348 & \textbf{35.7451} & \textbf{0.0753} & 31.2300 & 0.9137\\ 
 & DeblurGAN &  18.7211 & 0.0545 & 29.6118 &	0.9274 & 50.6455 &	0.1291 & 29.7493 & 0.8971  \\
 & Restormer & 21.9359 & 0.0523 & \textbf{34.1125} & \textbf{0.9664} & 51.3572 &	0.1043 & \textbf{32.7498} & \textbf{0.9377} \\ 
 & MPRNet & 23.0067 & 0.0544 & 34.0028 &	0.9654 & 64.2903 & 0.1176 & 32.5545 &	0.9352\\ 
 & NAFNet & 20.5896 & 0.0503 & 34.0764 & 0.9661 & 51.4628 & 0.1063 & 32.4929 & 0.9322\\ 
 & DnCNN & 26.8650 & 0.0726 & 33.1393 & 0.9542 & 111.3414 & 0.1628 & 31.8040 & 0.9217\\ 
 & SRAD & 47.7516 & 0.2374 & 27.5801 & 0.8476 & 70.5202 & 0.3565 & 27.7434 & 0.8318\\ 
  & CBM3D & 25.1978 & 0.1282 & 29.5931 & 0.9067 & 74.0214 & 0.2553 & 29.8244 & 0.8836\\
\hline
\multirow{5}{*}{0.08} & Ours (ODE) & \textbf{19.0333} & \textbf{0.0580} & 29.8902 & 0.9148 & \textbf{50.6311} & \textbf{0.1067} & 29.6971 & 0.8887\\ 
 & DeblurGAN & 30.8884 &	0.0737 & 26.0591 & 0.8690 & 79.0759 & 0.1668 & 27.6491 & 0.8553  \\
 & Restormer & 27.5327 & 0.0728 & \textbf{32.5138} & \textbf{0.9551} & 67.5050 &	0.1437 & \textbf{31.1680} & \textbf{0.9162}\\ 
 & MPRNet & 29.0643 & 0.0772 & 32.3741 & 0.9534 & 90.8632 & 0.1700 & 30.9350 &	0.9116\\ 
 & NAFNet & 25.3441 & 0.0693 & 32.4876 & 0.9546 & 71.6357 & 0.1403 & 30.9993 & 0.9131 \\ 
 & DnCNN & 27.7345 & 0.0834 & 31.5783 & 0.9426 & 126.2552 & 0.1867 & 30.2481 & 0.8978 \\ 
 & SRAD & 60.6093 & 0.3162 & 25.7904 & 0.7835 & 91.9534 & 0.4235 & 26.3913 & 0.7772\\ 
  & CBM3D & 31.2749 & 0.1774 & 27.4770 & 0.8726 & 89.7331 & 0.3181 & 28.1350 & 0.8433\\
\hline
\multirow{5}{*}{0.12} & Ours (ODE) & \textbf{24.3077} & \textbf{0.0774} & 28.5567 & 0.8994 & \textbf{63.2679} & \textbf{0.1333} & 28.8183 & 0.8727\\ 
 & DeblurGAN &  38.3176 & 0.0902 & 24.3920 &	0.8360  & 93.9357 &	0.1916 & 26.1810 & 0.8218\\
 & Restormer & 29.6475 & 0.0848 & \textbf{31.5912} & \textbf{0.9473} & 76.1931 & 0.1653 & \textbf{30.2727} & \textbf{0.9016}\\ 
 & MPRNet & 31.5731 & 0.0906 & 31.4240 &	0.9447 & 98.2022 & 0.1884 & 30.0573 &	0.8965\\ 
 & NAFNet & 27.0304 & 0.0805 & 31.5588 & 0.9466 & 77.2864 & 0.1644 & 30.0889 & 0.8977\\ 
 & DnCNN & 33.2722 & 0.1113 & 30.5007 & 0.9281 & 138.2030 & 0.2361 & 29.2134 & 0.8710\\ 
 & SRAD & 70.4303 & 0.3853 & 24.3354 & 0.7311 & 107.6315 & 0.4712 & 25.2329 & 0.7247\\  
 & CBM3D & 36.2575 & 0.2128 & 26.1749 & 0.8481 & 100.1406 & 0.3570 & 27.1042 & 0.8144\\ \hline
\end{tabular}
}
\caption{Comparison of different denoising methods for various metrics at different noise levels on CelebA and UC Merced Landuse dataset. The best performance for each metric is highlighted in bold. }
\label{tab:benchmarks}
\end{table*}

\subsection{Experiment settings}
We ran our experiments on CelebA \cite{liu2015faceattributes} and UC Merced Land Use \cite{yang2010bag} datasets, using U-Net \cite{ronneberger2015u} as the backbone architecture for our neural networks. The training was done on 100,000 images from the CelebA dataset, while testing was performed on 2,096 images from CelebA holdout set and another 2,096 images from UC Merced Land Use, images were resized to 224x224 pixels. We did not finetune on the land use dataset since we wanted to test the generalization of directly modeling the noise dynamics. 
% For finetuning, we provide a discussion using Doob's $h$-transform to speed up the process in Appendix.

For baseline models, we have selected several denoising methods, ranging from classical signal processing techniques to modern CNN-based frameworks:
\begin{itemize}
    \item The block-matching and 3D filtering (BM3D) was proposed in \cite{dabov2007image}, it partitions the image into multiple smaller patches and performs collaborative filtering to remove the noise. The method takes advantage of redundancy and consistent information across patches to generate a clean image, it achieved state-of-the-art performance at the times without requiring prior knowledge about noise statistics. 
    \item Speckle reducing anisotropic diffusion (SRAD), proposed in \cite{yu2002speckle}, is a method specifically designed for despeckling or multiplicative noise removal. It assumes a set of partial differential equations modeling the noise characteristics and iteratively adjusts the value of each pixel based on the local intensity variations.
    \item In \cite{zhang2017dncnn}, DnCNN was proposed to perform Gaussian denoising in latent space using CNN and residual learning. It popularized the idea of learning the noise (residuals) of a noisy image instead of directly learning the clean version. DnCNN remains an important tool today and is widely available in image processing toolboxes such as Matlab.
    \item MPRNet \cite{Zamir2021MPRNet} and Restormer \cite{Zamir2021Restormer}, a family of image restoration models that utilize multiple scales and transformer architecture to perform the inversion. These models are very competitive with NAFNet in terms of PSNR and SSIM, but we found that NAFNet still beats them in perception metrics.
    \item NAFNet \cite{chen2022simple}, the current state-of-the-art in noise removal and image restoration. It incorporates multiple advances made by DnCNN, HINet \cite{Chen_2021_CVPR}, SwinIR \cite{liang2021swinir}, BasicSR \cite{basicsr} and others. We make comparison with the width-64 version of NAFNet, since this model has 116M parameters, which is comparable to ours, which has 108M.
    \item Finally, since Generative Adversarial Networks (GANs) do optimize for perception metrics, we also make comparison with GAN-based models by adapting DeblurGAN \cite{DeblurGAN} to our dataset. 
\end{itemize} 

Regarding training settings, we trained our model with $T=500$ diffusion steps, linear noise schedule $\sigma(k)\in [0.0001, 0.02]$, and Adam optimizer \cite{kingma2014adam}. For DnCNN\footnote{\url{https://github.com/cszn/DnCNN}}, NAFNet\footnote{\url{https://github.com/megvii-research/NAFNet}}, MPRNet\footnote{\url{https://github.com/swz30/MPRNet}}, Restormer\footnote{\url{https://github.com/swz30/Restormer}}, and DeblurGAN\footnote{\url{https://github.com/KupynOrest/DeblurGAN}}, we followed the optimal training options recommended by the authors on Github. All models are trained for 100 epochs. Training was done on 2x RTX3090 under Ubuntu using Pytorch framework. Since DnCNN, NAFNet, MPRNet, Restormer, and DeblurGAN need to be trained for some specific noise levels, we chose three different noise variances $[0.04, 0.08, 0.12]$ for the noise term in \eqref{eq:mult_noise}, this corresponds to $T=100, 200, 300$ in our diffusion process formulation. We note that, while these models need to be trained for each noise level, our framework only needs to be trained once, then inference can be run at different noise levels since the noise dynamics is already captured by the SDE formulation. 

For evaluation metrics, we chose to evaluate on perception-based metrics FID, LPIPS and traditional signal processing metrics PSNR, SSIM. While PSNR and SSIM are important measurements in computer vision, they have been shown to not correlate well with human perception on image restoration tasks \cite{zhang2018perceptual}. In our experiments, we also observed substantial discrepancies between these two types of metrics, more detailed discussion is provided in the next section.

Our code is provided in the appendix.
\begin{figure*}\small
    \setlength{\tabcolsep}{1.5pt}
    \begin{tabular*}{0.92\textwidth}{@{\extracolsep{\fill}} ccccccccc }
    Original & Noisy & Ours &MPRNet & Restormer & NAFNet & DnCNN & SRAD & BM3D
    \end{tabular*}
  \centering
  % \tiny{Original Noisy Ours NAFNet DnCNN SRAD BM3D}
  % \includegraphics[width=0.45\textwidth]{fig/test_celeba_275_945_2046_1850.png}
  % \includegraphics[width=0.45\textwidth]{fig/main_paper_fig3.png}
  % \includegraphics[width=\textwidth]{fig/test_celeba_raw_275_945_2046_1850.png}
  % \includegraphics[width=\textwidth]{fig/test_celeba_raw_275_945_2046_1850_crop_30_40.png}
  \includegraphics[width=\textwidth]{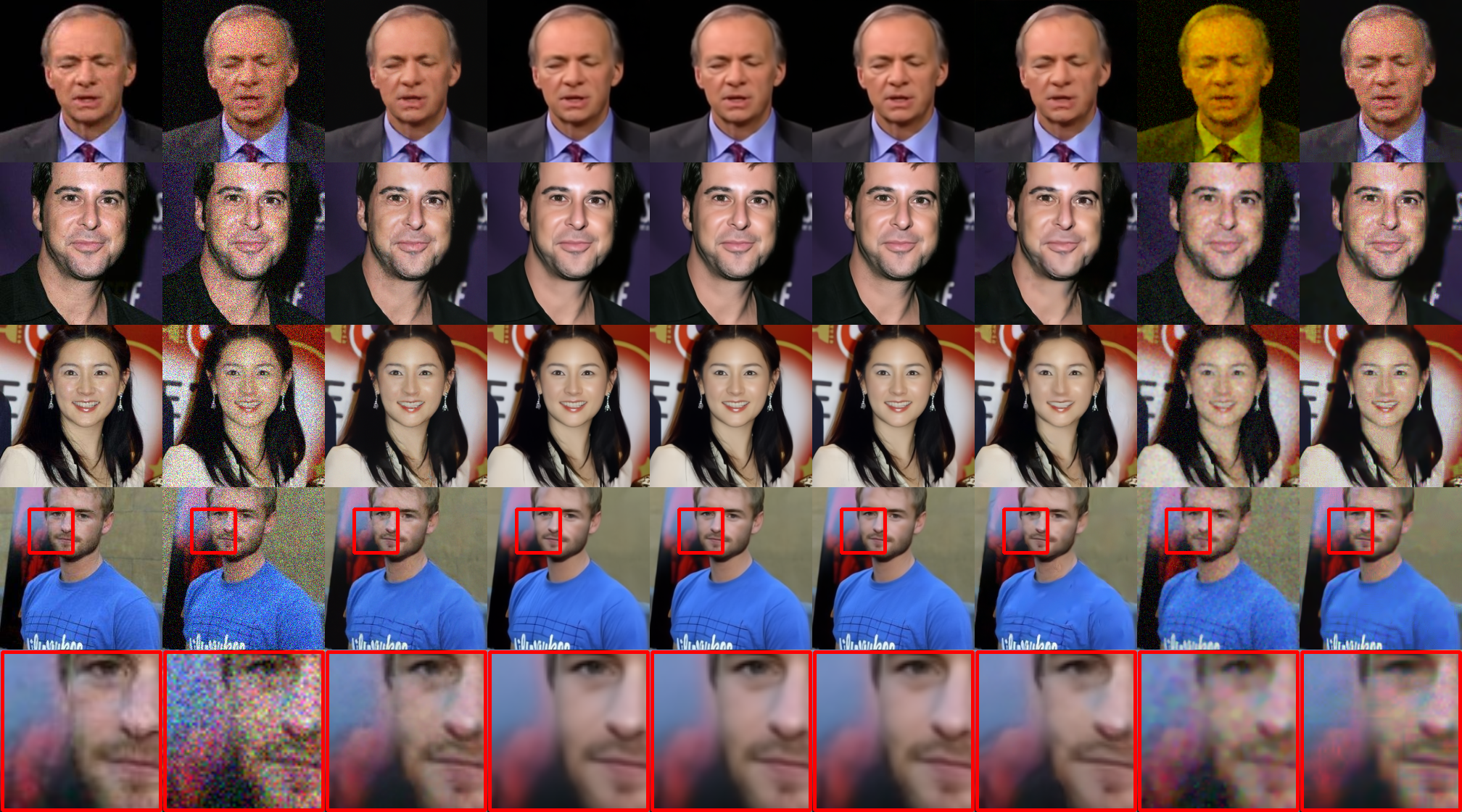}
  \caption{\small Comparing between different denoising models on randomly selected CelebA images, at noise level $0.08$. The first two columns include the original images and their noised versions, respectively. These are followed by the results generated by our method and other popular techniques. Last two rows present a zoomed in example. }\label{fig:compare_celeba}
\end{figure*}
\subsection{Experiment results}

\textbf{Sampling techniques.}
We evaluate our model with three different sampling techniques: stochastic sampling \eqref{eq:discrete_reverse}, ODE sampling \eqref{eq:discrete_ode_reverse}, and DDIM sampling \eqref{eq:ddim_sampling}, on images from CelebA holdout set, generated from noise level $0.12$. Results are presented in Table \ref{tab:sampling_celeba}. It can be seen that samples generated using ODE technique perform substantially better across all metrics, especially in FID and LPIPS. Compared to stochastic sampling, deterministic samplers are up to 57\% better at FID, and 66\% better at LPIPS. This results in observable differences in generated samples, as can be seen from the qualitative examples provided in Figure \ref{fig:quali_sampling_methods}. 

\textbf{Results on CelebA dataset.} Quantitative results are presented in the left half of Table \ref{tab:benchmarks}. Compared to the state-of-the-art models NAFNet and Restormer, our method shows slightly worse performance in "per-pixel" metrics, while achieving significantly better FID and LPIPS scores, across all tested noise levels. This is also true when comparing to DnCNN and MPRNet. We note that these models were architected to directly optimize PSNR during training, thus they strive to achieve the best fidelity at the cost of diverging from the input distribution. For classical methods, CBM3D (the RGB version of BM3D) performs respectably, sometimes even coming close to DnCNN. In contrast, SRAD falls far behind in all metrics, we suspect this is because the tests were conducted using RGB images, while this method was originally designed for grayscale samples only. 

For DeblurGAN, while it performs respectably at low noise level, it gets substantially worse than other methods across all metrics when more noise is present in the images. In either case, our method also beats it decisively. This is expected since GANs are difficult to train optimally, and are often beaten by Diffusion models in quality of generated samples \cite{dhariwal2021diffusion}.  

These observations can also be seen in Figure \ref{fig:compare_celeba}, where we present denoising results of theses models at noise level $0.08$. Compared to our method, other techniques suffer from over-smoothing and they usually generate samples that lack high-frequency details. This is the drawback of using PSNR as an optimization objective, where the denoisers have the tendency of collapsing into the mean statistics of the images, creating smoothing effect. In contrast, our proposed model tends to be much better at preserving finer details, such as facial hair and clothes wrinkles.
\begin{figure*}
  % \tiny{Original Noisy Ours NAFNet DnCNN SRAD BM3D}
  % \includegraphics[width=0.45\textwidth]{fig/test_landuse_44_1982_167_1840.png}
  % \includegraphics[width=0.45\textwidth]{fig/main_paper_fig4.png}
  \small
    \setlength{\tabcolsep}{1.5pt}
    \begin{tabular*}{0.9\textwidth}{@{\extracolsep{\fill}} ccccccccc }
    Original & Noisy & Ours &MPRNet & Restormer & NAFNet & DnCNN & SRAD & BM3D
    \end{tabular*}
  \centering
  % \tiny{Original Noisy Ours NAFNet DnCNN SRAD BM3D}
  % \includegraphics[width=0.45\textwidth]{fig/test_celeba_275_945_2046_1850.png}
  % \includegraphics[width=0.45\textwidth]{fig/main_paper_fig3.png}
  % \includegraphics[width=\textwidth]{fig/test_celeba_raw_275_945_2046_1850.png}
  % \includegraphics[width=\textwidth]{fig/test_raw_landuse_zoom_1982_crop_40_90_167_100_60.png}
  \includegraphics[width=\textwidth]{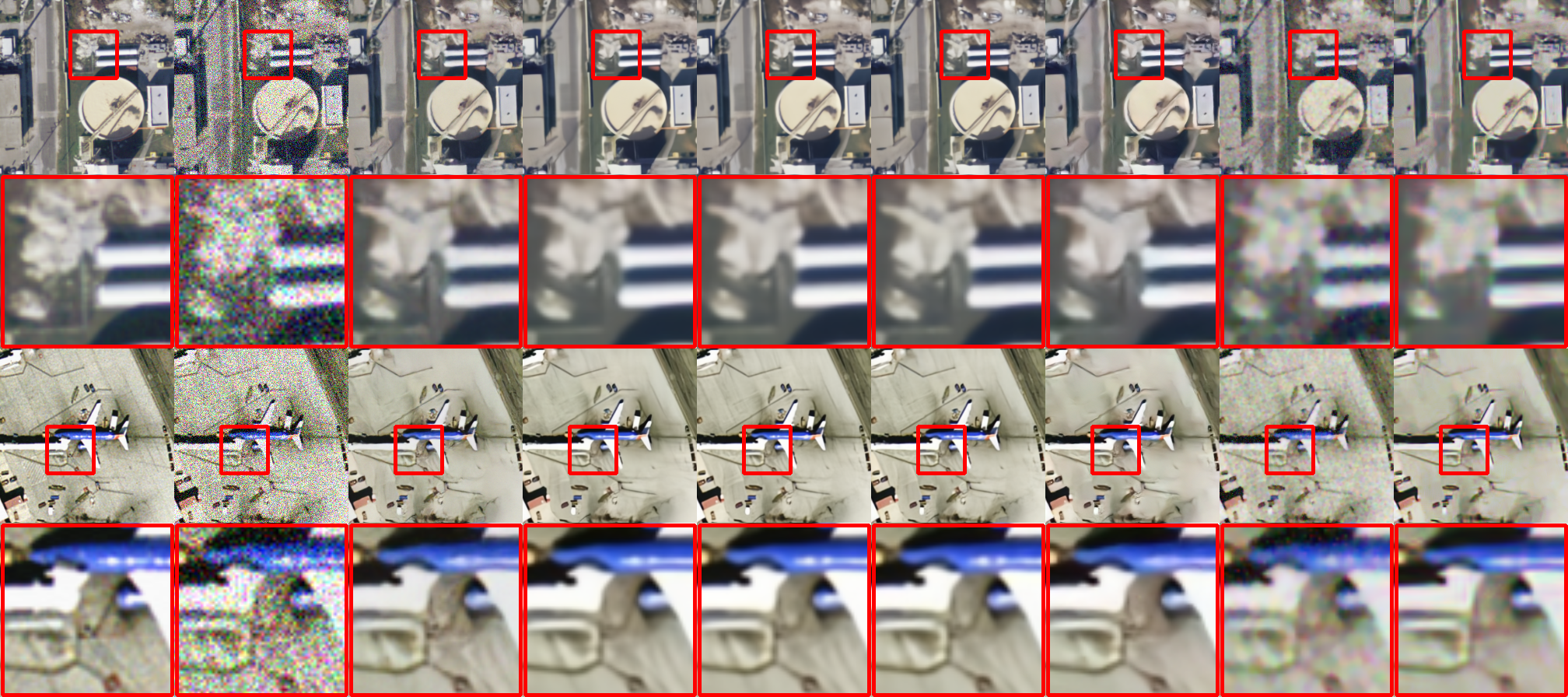}
  \caption{\small Comparing between different denoising models on randomly selected UC Merced Land Use images, at noise level $0.12$. The first two columns include the original images and their noised versions, respectively. These are followed by the results generated by our method and other popular techniques. }\label{fig:compare_landuse}
\end{figure*}

\textbf{Out-of-distribution results on Land Use dataset.} To evaluate the generalization capability of our method, we re-run the previous experiments on the UC Merced Land Use dataset, which is a small dataset (containing 2,096 samples) captured from satellites. Besides being out-of-distribution, we note that this is a much harder dataset to do denoising on, due to the blurriness and color shifting of satellite imaging. We present quantitative results in the right half of Table \ref{tab:benchmarks}. While there is large degradation in performance across the board, we again see that our method achieves the best FID and LPIPS scores, while being slightly worse in terms of PSNR and SSIM. Surprisingly, DnCNN performs much worse in terms of FID, while remaining competitive in the other metrics. This could be related to the incorrect assumption of the FID calculation, which was recently discussed in \cite{jayasumana2024rethinking}.

Qualitatively, from Figure \ref{fig:compare_landuse}, it is observed that our method retains finer details of the images. DnCNN, NAFNet, MPRNet, Restormer, and DeblurGAN produce over-smoothed samples, losing details of the ground, while SRAD and BM3D generate images that completely lack high-frequency components.

Overall, we see that our method achieves competitive performance on PSNR and SSIM, while producing more realistic samples that are closer to the true input distribution, as measured by FID and LPIPS metrics. Furthermore, we show empirically that the method can generalize well to out-of-distribution dataset, which could be crucial in applications where data samples are limited. We provide more qualitative examples in the Appendix.
\section{Conclusion}

In conclusion, this paper introduces a novel SDE-based diffusion model for removing multiplicative noise. The work presents the construction of the forward and reverse SDEs that directly captures the dynamics of the noise model. In addition, it also establishes the training objective as well as multiple different sampling equations based on Probability flows and DDIM techniques. The proposed model is compared to classical image processing algorithms, including BM3D and SRAD, as well as the modern CNN-based methods. Extensive experiments on different datasets demonstrate that our method outperforms the current state-of-the-art denoising models in perception-based metrics across all noise levels, while still remaining competitive in PSNR and SSIM.

Going forward, we will explore the application of diffusion steps reduction techniques to noise removal problems. While these techniques have been applied successfully in generative tasks, it greatly reduces the quality of generated samples in our problem. Thus, care need to be taken when dealing with tasks that are sensitive to small perturbation like denoising. Furthermore, while deterministic sampling is usually used to speedup generation, it is a desirable property in noise removal tasks. Specifically, we would like the process to produce the exact clean image, not something close in terms of distribution, which is modeled by the current diffusion loss. This has connections to conditioning using Doob's $h$-transform, and could hold interesting research venue, we leave it to future works. 
\vspace{-1mm}
{
    \small
    \bibliographystyle{ieeenat_fullname}
    \bibliography{main}
}

\onecolumn
\clearpage
\setcounter{section}{0}
\setcounter{page}{1}
\renewcommand\thesection{\Alph{section}}

\maketitlesupplementary

\newtheorem{theorem}{Theorem}
\newtheorem{definition}{Definition}
\newtheorem{example}{Example}
\newtheorem{lemma}{Lemma}
\newtheorem*{lemma*}{Lemma}
 \newcommand{\ind}{\perp\!\!\!\!\perp} 

\section{Solutions to the forward SDE}
Consider the 1-dimensional SDE
\begin{equation}
    dx=\alpha(t) xd\beta(t)\label{eq:forward_sed}
\end{equation}
This can be solved by applying Itô's formula to $\log x$
\begin{align}
    d\log x &= \frac{1}{x}dx -\frac{1}{2x^{2}}(dx)^{2}\\
    &=\frac{1}{x}\alpha(t) xd\beta(t)-\frac{1}{2x^{2}}\alpha^{2} x^{2}(d\beta(t))^{2}\\
    &=\alpha(t) d\beta(t) -\frac{1}{2}\alpha^{2}(t)dt \\
    \Rightarrow\;\; \log x_{t} &= \log x_{0} -\int_{0}^{t}\frac{1}{2}\alpha^{2}(\tau)d\tau + \int_{0}^{t}\alpha(\tau)d\beta(\tau)\label{eq:forward_sde_log}
\end{align}
Recalling a well-known lemma
\begin{lemma}\label{lemma:1.2.1.deter_int}
    Let $f(t)$ be some function that is square-integrable, i.e. $\int_{0}^{t}f^{2}(s)ds <\infty$, and $\beta(t)$ be a some Brownian motion, then
    \begin{equation}
        \int_{0}^{t}f(s)d\beta \sim \mathcal{N}\Big(0, \int_{0}^{t}f^{2}(s)ds\Big)
    \end{equation}
    Proof. Since Riemann's sum of $\int_{0}^{t}f(s)d\beta$ exists if we fix the midpoints, let $t_{k}=\frac{k}{2^{n}}t$, then
    \begin{align}
        \int_{0}^{t}f(s)d\beta &= \lim_{n\rightarrow\infty}\sum_{k=0}^{2^{n}-1}f(t_{k})(\beta_{t_{k+1}}-\beta_{t_{k}})
    \end{align}
    and since increment of Brownian motion follows $\mathcal{N}(0,\Delta t)$, where $\Delta t = 2^{-n}t$, thus
    \begin{align}
        \sum_{k=0}^{2^{n}-1}f(t_{k})(\beta_{t_{k+1}}-\beta_{t_{k}}) \sim \mathcal{N}\Big(0, \sum_{k=0}^{2^{n}-1}f^{2}(t_{k})2^{-n}t\Big)
    \end{align}
    Now we can take the limit
    \begin{align}
        \lim_{n\rightarrow\infty} \sum_{k=0}^{2^{n}-1}f^{2}(t_{k})2^{-n}t &= \lim_{n\rightarrow\infty} \sum_{k=0}^{2^{n}-1}f^{2}(t_{k})\Delta t\\
        &=\int_{0}^{t}f^{2}(s)ds
    \end{align}
    this completes the proof.
\end{lemma}
Applying this lemma to \eqref{eq:forward_sde_log} gives
\begin{align}
    \log x_{t}&= \log x_{0} -\int_{0}^{t}\frac{1}{2}\alpha^{2}(\tau)d\tau + \Big(\int_{0}^{t}\alpha^{2}(\tau)d\tau\Big)^{\frac{1}{2}}n \nonumber\\
    &\;\;\;\;\;\;\;\;\;\;\;\;\;\;\;\;n\sim\mathcal{N}(0,1) \nonumber
\end{align}
Taking exponential of the last equation yields the desired result.

\section{Derivations of the reverse SDEs}
\subsection{In pixel domain}
For the reverse SDE, we need to use the more general form of Anderson's theorem
\begin{align}
    dx_{T-t} &= \Big( -f(x,T-t)+ \frac{1}{p_{T-t}(x_{T-t})}\nabla L^{2}(x, T-t) p_{T-t}(x_{T-t})\Big)dt \nonumber+ L(x, T-t)d\beta_{T-t}
\end{align}
Which in our case simplifies to
\begin{align}
    dx_{T-t} &= \alpha^{2}(T-t)\frac{1}{p_{T-t}(x_{T-t})}\nabla x^{2}_{T-t}p_{T-t}(x_{T-t})dt + \alpha(T-t) x_{T-t}d\beta_{T-t} \nonumber\\
    &=\alpha^{2}(T-t)\Big( 2x_{T-t} + x^{2}_{T-t}\nabla\log p_{T-t}(x_{T-t}) \Big)dt + \alpha(T-t) x_{T-t}d\beta_{T-t}
\end{align}
This reverse SDE is more complicated to work with because of the spatial dependency of the noise term. Due to this, in order to achieve the same convergence guarantee as Euler-Maruyama discretization, more sophisticated schemes such as Milstein's correction needs to be employed \cite{kloeden1992stochastic}, making the generative process much more involved. This motivates us to use the logarithmic formulation, which is shown next.  

\subsection{In logarithmic domain}
Recall that if we apply the logarithmic transform $y=\log x$, then the forward SDE takes the form
\begin{equation}
     dy_{t}= -\frac{1}{2}\alpha^{2}(t)dt + \alpha(t)d\beta(t)
\end{equation}
which is a simple Wiener process. Applying the previously mentioned Anderson's theorem gives
\begin{align}
    dy_{T-t} &= \frac{1}{2}\alpha^{2}(T-t)dt+ \alpha^{2}(T-t)\frac{1}{p_{T-t}(y_{T-t})}\nabla p_{T-t}(y_{T-t})dt  \alpha(T-t) d\beta_{T-t} \nonumber\\
    &=\Big(\frac{1}{2}\alpha^{2}(T-t)+ \alpha^{2}(T-t)\nabla \log p_{T-t}(y_{T-t})\Big)dt + \alpha(T-t) d\beta_{T-t}
\end{align}
Since the corruption applies to each pixel independently, this can be trivially extended to multivariate case to obtain the result mentioned in the paper.

\section{Derivation of the deterministic sampling equation using Implicit models}
Recall the discretized forward equation presented in the main paper
\begin{align}
    \mathbf{y}_{k} &= \mathbf{y}_{k-1} -\frac{1}{2}\big(\sigma(k)-\sigma(k-1)\big)\mathbf{1} + \sqrt{\sigma(k)-\sigma(k-1)}\mathbf{n}_{k}\nonumber\\
    &= \mathbf{y}_{0} -\frac{1}{2}\big(\sigma(k)-\sigma(0)\big)\mathbf{1} + \sqrt{\sigma(k)-\sigma(0)}\mathbf{n}_{k}\label{eq:discrete_forward}
\end{align}
This has the Gaussian transition kernel 
\begin{align}
    p(\mathbf{y}_{k}|\mathbf{y}_{k-1})&= \mathcal{N}\Big(\mathbf{y}_{k-1}-\frac{1}{2}\big(\sigma(k)-\sigma(k-1)\big)\mathbf{1}, \big(\sigma(k)-\sigma(k-1)\big)\mathbf{I}\Big)\\
    p(\mathbf{y}_{k}|\mathbf{y}_{0})&= \mathcal{N}\Big(\mathbf{y}_{0}-\frac{1}{2}\big(\sigma(k)-\sigma(0)\big)\mathbf{1}, \big(\sigma(k)-\sigma(0)\big)\mathbf{I}\Big)
\end{align}
Consider the non-Markovian kernel
\begin{align}
    q&(\mathbf{y}_{k-1}|\mathbf{y}_{k}, \mathbf{y}_{0}) = \mathcal{N}(\bm{\mu}_{k}, \bm{\Sigma}_{k})\\
    &\bm{\mu}_{k}= \mathbf{y}_{0}-\frac{1}{2}\eta(k-1)\mathbf{1} + \frac{\sqrt{\eta(k-1)-\zeta_{k}^{2}}}{\sqrt{\eta(k)}}\big(\mathbf{y}_{k}-\mathbf{y}_{0}+\frac{1}{2}\eta(k)\mathbf{1}\big)\\
    &\bm{\Sigma}_{k}=\zeta_{k}^{2}\mathbf{I}
\end{align}
where $\zeta_{k}^{2}$ is a new parameter controlling the variance of the process, and $\eta(k)=\sigma(k)-\sigma(0)$. We now show that $q(\mathbf{y}_{k-1}|\mathbf{y}_{0})$ matches $p(\mathbf{y}_{k-1}|\mathbf{y}_{0})$. From \cite{bishop2006pattern} (2.115), $q(\mathbf{y}_{k-1}|\mathbf{y}_{0})$ is Gaussian $\mathcal{N}(\bm{\mu}_{k-1}, \bm{\Sigma}_{k-1})$, and has the following forms
\begin{align}
    \bm{\mu}_{k-1} &= \mathbf{y}_{0}-\frac{1}{2}\eta(k-1)\mathbf{1} + \frac{\sqrt{\eta(k-1)-\zeta_{k}^{2}}}{\sqrt{\eta(k)}}\big(\textcolor{blue}{\mathbf{y}_{0}-\frac{1}{2}\eta(k)\mathbf{1}}-\mathbf{y}_{0}+\frac{1}{2}\eta(k)\mathbf{1}\big)\nonumber\\
    &= \mathbf{y}_{0}-\frac{1}{2}\eta(k-1)\mathbf{1}\\
    \bm{\Sigma}_{k-1} &= \zeta_{k}^{2}\mathbf{I} + \Big(\frac{\sqrt{\eta(k-1)-\zeta_{k}^{2}}}{\sqrt{\eta(k)}}\Big)^{2}\eta(k)\mathbf{I}\nonumber\\
    &= \zeta_{k}^{2}\mathbf{I} + \big(\eta(k-1)-\zeta_{k}^{2}\big)\mathbf{I}\nonumber\\
    &= \eta(k-1)\mathbf{I}
\end{align}

Thus, $q(\mathbf{y}_{k-1}|\mathbf{y}_{0})$ has the Gaussian kernel
\begin{align}
    q(\mathbf{y}_{k-1}|\mathbf{y}_{0}) &= \mathcal{N}(\bm{\mu}_{k-1}, \bm{\Sigma}_{k-1}) \nonumber\\
    &= \mathcal{N}\Big( \mathbf{y}_{0}-\frac{1}{2}\eta(k-1)\mathbf{1},  \eta(k-1)\mathbf{I}\Big) \nonumber\\
    &= \mathcal{N}\Big(\mathbf{y}_{0}-\frac{1}{2}\big(\sigma(k-1)-\sigma(0)\big)\mathbf{1},\big(\sigma(k-1)-\sigma(0)\big)\mathbf{I}\Big) \nonumber\\
    &= p(\mathbf{y}_{k-1}|\mathbf{y}_{0})
\end{align}
This completes the proof.
\section{Additional experiment results}
\begin{figure}
    \small
    \setlength{\tabcolsep}{1.5pt}
    \begin{tabular*}{0.92\textwidth}{@{\extracolsep{\fill}} ccccccccc }
    Original & Noisy & Ours &MPRNet & Restormer & NAFNet & DnCNN & SRAD & BM3D
    \end{tabular*}
  \centering
  % \tiny{Original Noisy Ours NAFNet DnCNN SRAD BM3D}
  \includegraphics[width=\textwidth]{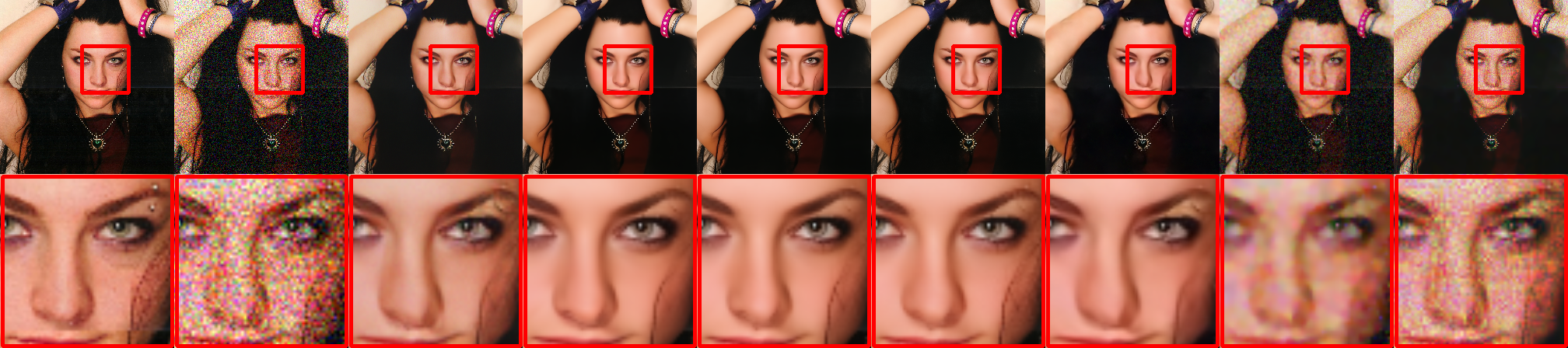}
  \includegraphics[width=\textwidth]{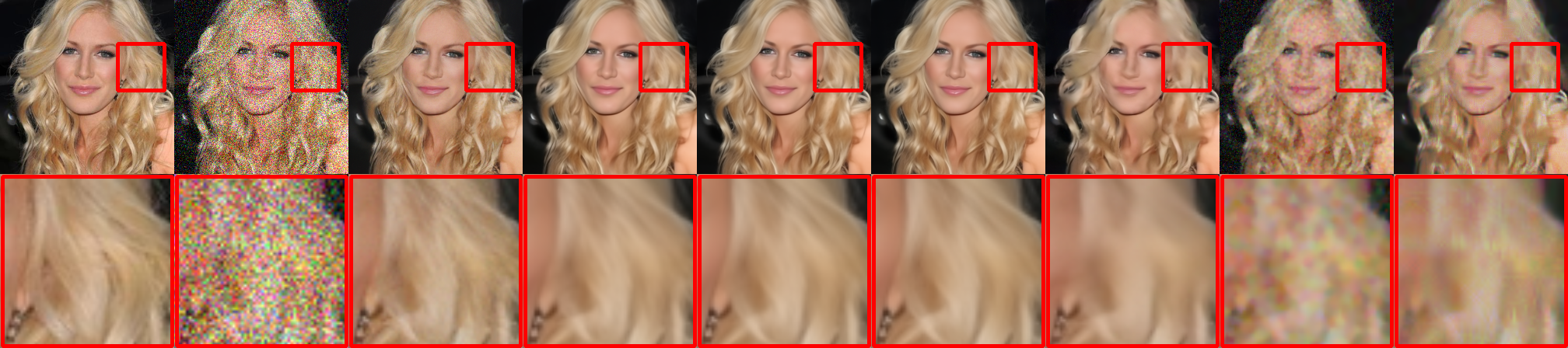}
  \includegraphics[width=\textwidth]{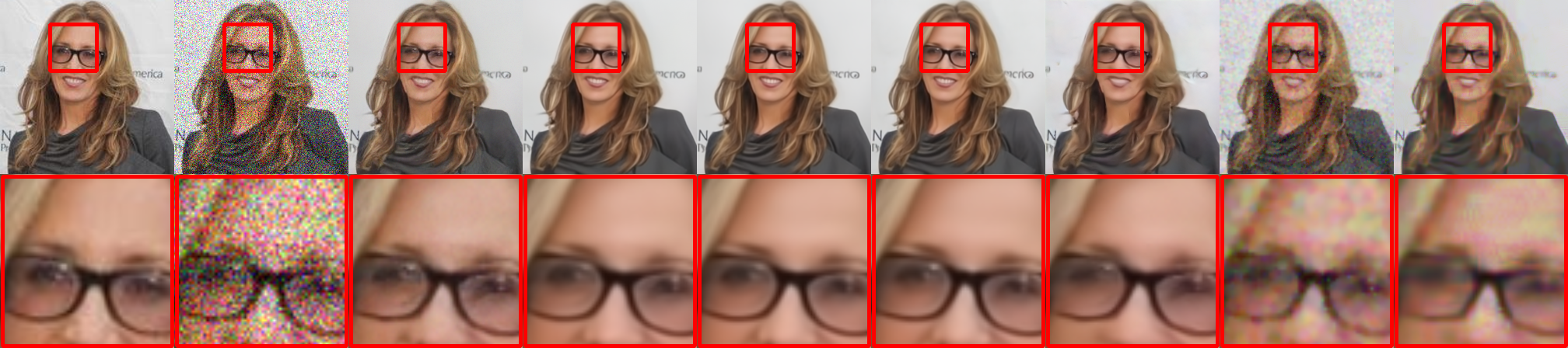}
  \includegraphics[width=\textwidth]{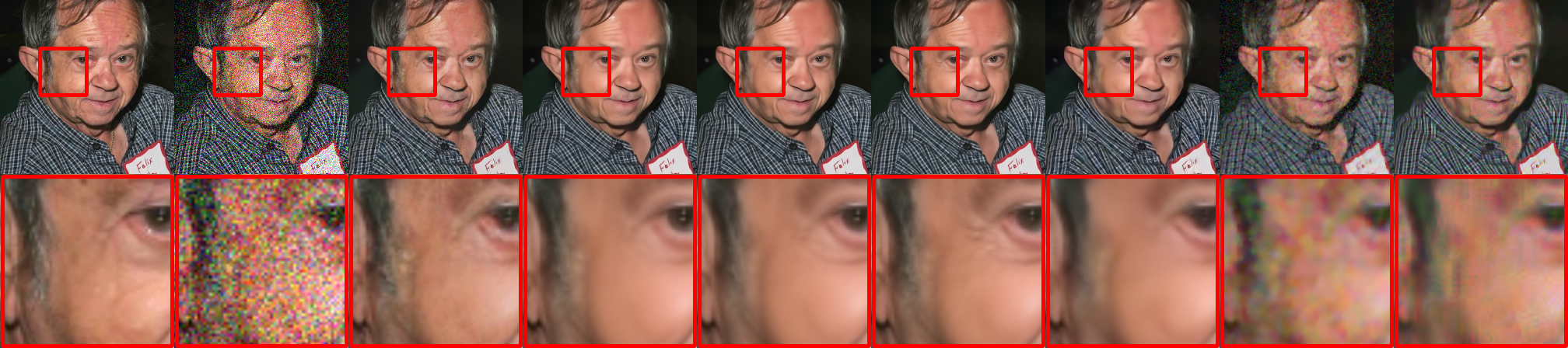}
  \caption{\small Comparing between different denoising models on randomly selected CelebA images, at noise level $0.12$. The first two columns include the original images and their noised versions, respectively. These are followed by the results generated by our method and other popular techniques. }\label{fig:compare_celeba_app}
\end{figure}

\begin{figure}
    \small
    \setlength{\tabcolsep}{1.5pt}
    \begin{tabular*}{0.92\textwidth}{@{\extracolsep{\fill}} ccccccccc }
    Original & Noisy & Ours &MPRNet & Restormer & NAFNet & DnCNN & SRAD & BM3D
    \end{tabular*}
  \centering
  % \tiny{Original Noisy Ours NAFNet DnCNN SRAD BM3D}
  \includegraphics[width=\textwidth]{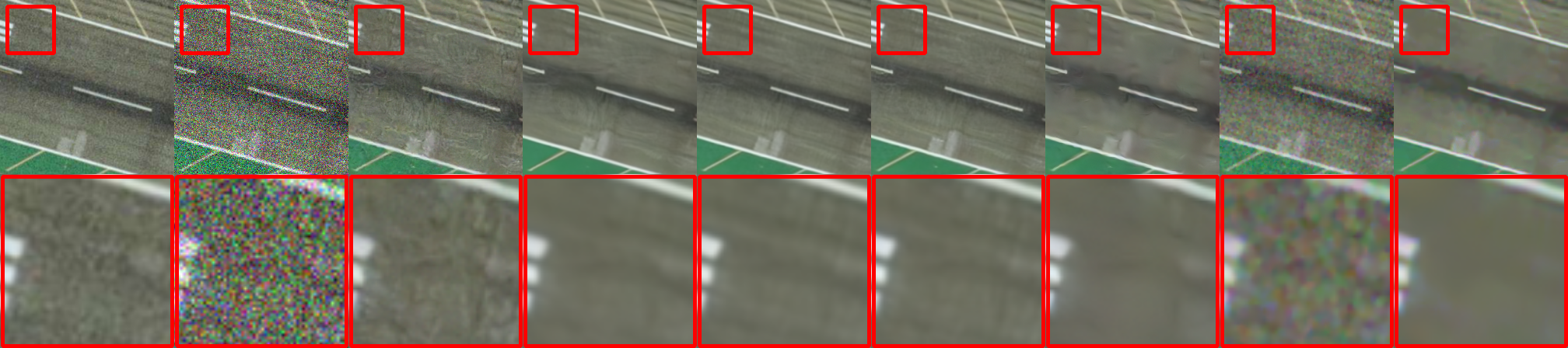}
  \includegraphics[width=\textwidth]{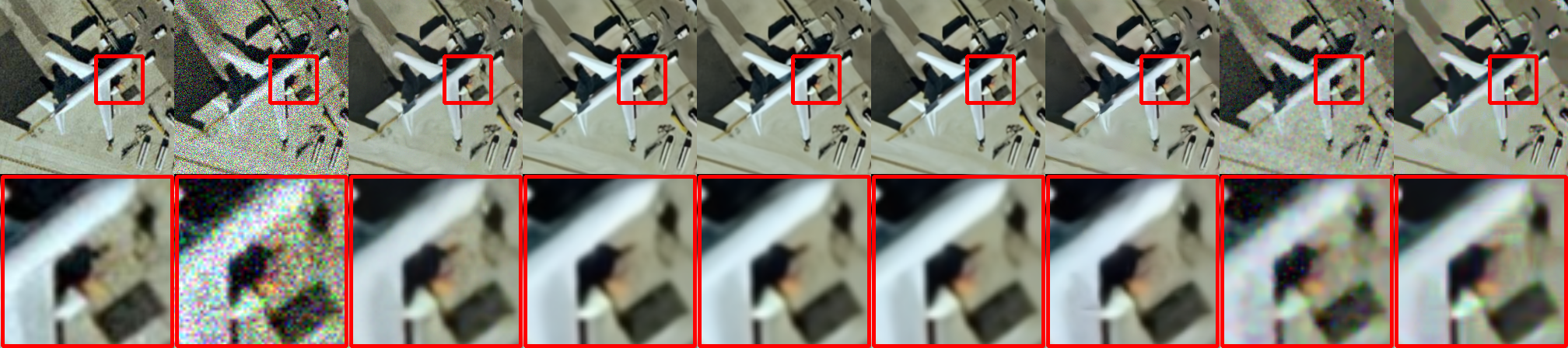}
  \includegraphics[width=\textwidth]{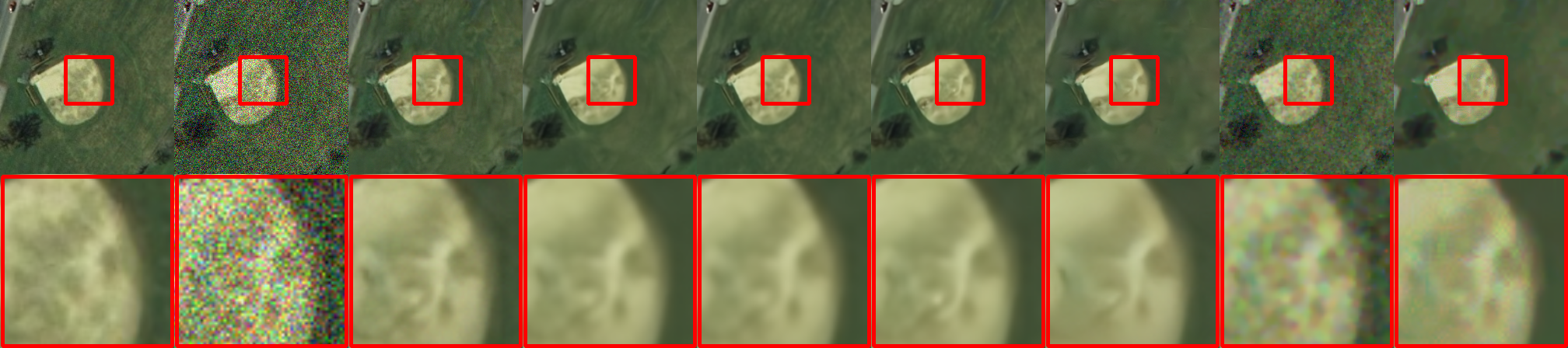}
  \includegraphics[width=\textwidth]{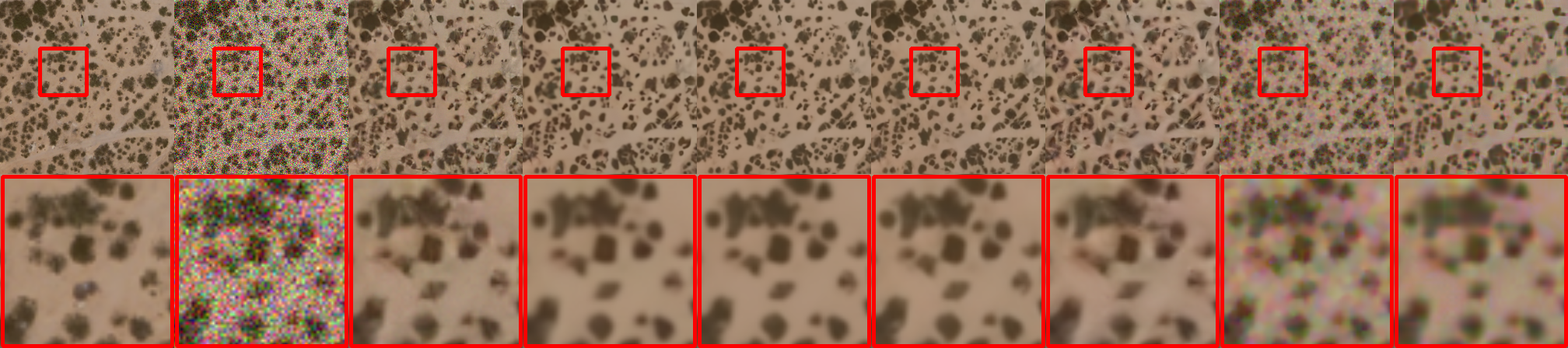}
  \caption{\small Comparing between different denoising models on randomly selected UC Merced Land Use images, at noise level $0.12$. The first two columns include the original images and their noised versions, respectively. These are followed by the results generated by our method and other popular techniques. }\label{fig:compare_landuse_app}
\end{figure}

We provide more qualitative test samples in Figures \ref{fig:compare_celeba_app} and \ref{fig:compare_landuse_app}. In Fig.\ref{fig:compare_celeba_app}, high-frequency components such as hair and facial wrinkles are better preserved by our method. Whereas other methods have a strong tendency to oversmooth thing, and produce samples that are lacking in this regard. Similarly, in Fig.\ref{fig:compare_landuse_app}, our method tends to retain more high-frequency components compared to others. This results in images that, perceptually, are closer to the original ones.

\end{document}